\documentstyle[12pt]{article}

\begin{document}

\vspace{2mm}

\vspace{2ex}

\begin{flushright}
Preprint MRI-PHY/96-24 \\

hep-th/9608026
\end{flushright}

\vspace{2ex}

\begin{center}
{\large \bf  Singularity Free (Homogeneous Isotropic) 

\vspace{2ex} 

             Universe in Graviton-Dilaton Models } \\

\vspace{6mm}
{\large  S. Kalyana Rama}
\vspace{3mm}

Mehta Research Institute, 10 Kasturba Gandhi Marg, 

Allahabad 211 002, India. 

\vspace{1ex}
email: krama@mri.ernet.in \\ 
\end{center}

\vspace{4mm}

\begin{quote}
ABSTRACT. We present a class of graviton-dilaton models 
in which a homogeneous isotropic universe, such as our 
observed one, evolves with no singularity at any time. 
Such models may stand on their own as interesting models 
for singularity free cosmology, and may be studied further 
accordingly. They may also arise from string theory. 
We discuss critically a few such possibilities. 
\end{quote}


\newpage

{\bf 1.} 
General Relativity (GR) is a beautiful theory and, more 
importantly, has been consistently successful in describing 
the observed universe. But, GR leads inevitably to 
singularities. In particular, our universe according to GR 
starts from a big bang singularity. We do not, however, 
understand the physics near the singularity nor know its 
resolution. The inevitability and the lack of understanding 
of the singularities point to a lacuna in our fundamental 
understanding of gravity itself. By the same token, 
a successful resolution of the singularities 
is likely to provide deep insights into gravity. 

In order to resolve the singularities, it is necessary 
to go beyond GR, perhaps to a quantum theory of gravity. 
A leading candidate for such a theory is string theory. 
However, despite enormous progress \cite{rev}, string 
theory has not yet resolved the singularities \cite{vafa}. 

Perhaps, one may have to await further progress in 
string theory, or a quantum theory of gravity, to be able 
to resolve singularities. However, there is still an avenue 
left open - resolve the singularities, if possible, within 
the context of a generalised Brans-Dicke model, and then 
enquire whether such a model can arise from string theory, 
or a quantum theory of gravity. 

Brans-Dicke theory is special for more than one 
reason. It appears naturally in supergravity 
and Kaluza-Klein theories, and also in all the known 
effective string actions. It is also perhaps the most 
natural extension of GR \cite{will}, which may explain 
its ubiquitous appearance in fundamental theories: Dicke 
had discovered long ago the lack of observational evidence 
for the principle of strong equivalence in GR. Together 
with Brans, he then constructed the Brans-Dicke theory 
\cite{bd} which obeys all the principles of GR except 
that of strong equivalence. 

Brans-Dicke theory contains a graviton, a scalar (dilaton) 
coupled non minimally to the graviton, 
and a constant parameter $\omega$. GR is obtained when 
$\omega = \infty$. In the generalised Brans-Dicke theory, 
commonly referred to as graviton-dilaton theory, $\omega$ 
is an arbitrary function of the dilaton \cite{will,gbd}. 
The graviton-dilaton theory contains an infinite number 
of models, one for every function $\omega$, very likely 
including all the effective actions that may arise 
from string theory, or a quantum theory of gravity. 

This perspective opens up an avenue in resolving 
the singularities: to find, if possible, the class of 
graviton-dilaton models which are singularity free. One 
may then study whether they can arise from string theory, 
or a quantum theory of gravity. On the other hand, such 
models may stand on their own as interesting models for 
singularity free cosmology and one may then study their 
implications in other cosmological and astrophysical 
contexts, which are likely to be novel \cite{k}. 
Such studies are fruitful and are likely to provide 
valuable insights. 

Accordingly, in this letter, we present a class of 
graviton-dilaton models where the function $\omega$ 
satisfies certain constraints. We show that a homogeneous 
isotropic universe, such as our observed one, evolves 
in these models with no big bang or any other 
singularity. The time continues indefinitely into 
the past and the future, without encountering any 
singularity. These models also satisfy the observational 
constraints imposed by solar system experiments. We then 
discuss critically a few possibilities of such a model 
arising from string theory. 


{\bf 2.} 
Consider the following graviton-dilaton action 
\begin{equation}\label{act}
S = \frac{1}{16 \pi G_N} \int d^4 x \sqrt{- g} \left( 
- \chi R + \frac{\omega (\chi)}{\chi} \; (\nabla \chi)^2 
\right) + S_M ({\cal M}, \; g_{\mu \nu})  \; , 
\end{equation}
where $G_N$ is Newton's constant, $\chi \ge 0$ is 
the dilaton, $\omega (\chi)$ is an arbitrary function 
characterising the theory, and $S_M$ is the action for 
matter fields, denoted collectively by ${\cal M}$ and 
coupling minimally to $g_{\mu \nu}$. The ``matter'' is 
assumed to be a perfect fluid with density $\rho$ 
and pressure $p$, related by $p = \gamma \rho$ 
where $- 1 \le \gamma \le 1$. The value of $\gamma$ 
indicates the nature of the ``matter''. For example, 
$\gamma = 0, \; \frac{1}{3}$, or $1$ indicates that 
the ``matter'' is dust, radiation, or massless scalar 
field respectively. If $\chi$ has a finite range 
$0 \le \chi \le \chi_{{\rm max}} < \infty$, as is 
the case in the present model, then the factor 
$\chi_{{\rm max}}$ can be absorbed into $G_N$ 
and the range of $\chi$ can be set to be 
$0 \le \chi \le 1$. Also, we will work in the units 
where $G_N = \frac{\chi_{{\rm max}}}{16 \pi}$. 

In the model we propose, derived originally in 
a different context in \cite{k}, $\chi$ has a finite 
range, set to be $0 \le \chi \le 1$, and the function 
$\Omega (\chi) \equiv 2 \omega (\chi) + 3$ is required 
to satisfy the following constraints only \cite{ein}: 
\begin{eqnarray}
{\bf {\rm (i)}}  & & 
\Omega (\chi) > 0, \; \; \; \; 0 \le \chi \le 1 
\; \; \; \;  
{\bf {\rm (ii)}}  \; \;  
\Omega (0) = \Omega_0 \le \frac{1}{3}  \nonumber \\ 
{\bf {\rm (iii)}}  & & 
\lim_{\chi \to 1} \Omega = 
\Omega_1 (1 - \chi)^{- 2 \alpha} \; , \; \; \; \; 
\frac{1}{2} \le \alpha < 1 \nonumber \\ 
{\bf {\rm (iv)}}  & & \frac{d^n \Omega}{d \chi^n} = 
{\rm finite} \; \; \; \; \forall \; n \ge 1, 
\; \; \; \; 0 \le \chi < 1  \; . \label{123} 
\end{eqnarray} 
$\Omega_0 > 0$ and $\Omega_1 > 0$ above are constants. 
For the purpose of illustration here, we further take 
$\Omega (\chi)$ to be a strictly increasing function. 
However, this constraint is not necessary. In fact, 
as will be shown elsewhere, the constraints (\ref{123}) 
can also be relaxed further without affecting the results. 

Consider a homogeneous isotropic universe. Then, 
$\chi = \chi(t)$ and the line element is given by 
$d s^2 = - d t^2 + e^{2 A(t)} \; (d r^2 + r^2 d S_2^2)$, 
where $d S_2$ is the line element on an unit sphere, and 
the curvature index $k$ is set to zero. (The results, 
however, are valid for $k = \pm 1$ also.) The equations 
of motion following from the action (\ref{act}) 
reduce to (see \cite{lessner}) 
\begin{eqnarray} 
\dot{A} + \frac{\dot{\chi}}{2 \chi} & = & 
\sqrt{\frac{\rho}{6 \chi} + \frac{\Omega \dot{\chi}^2} 
{12 \chi^2}} \label{ad} \\ 
\ddot{\chi} + 3 \dot{A} \dot{\chi} 
+ \frac{\dot{\Omega} \dot{\chi}}{2 \Omega} 
& = & \frac{(1 - 3 \gamma) \rho}{2 \Omega} \label{chidd} \\ 
\rho & = & \rho_0 e^{- 3 (1 + \gamma) A} \; , \label{rho} 
\end{eqnarray} 
where $p = \gamma \rho$ is used, upper dots denote 
$t$-derivatives, and $\rho_0 \ge 0$ is a constant. 
The square roots are to be taken always with 
a positive sign. 

Integrating equation (\ref{chidd}) once, we get 
\begin{equation}\label{chid} 
\dot{\chi} (t) = \frac{e^{- 3 A}}{\sqrt{\Omega}} \; 
(\sigma (t) + c) \; , \; \; \; \; 
\sigma (t) \equiv \frac{(1 - 3 \gamma) \rho_0}{2} 
\int^t_{t_i} dt \frac{e^{- 3 \gamma A}}{\sqrt{\Omega}} 
\end{equation}
and $c$ is a constant. Dividing (\ref{ad}) by 
$\frac{\dot{\chi}}{\chi}$ and substituting (\ref{chid}) 
for $\dot{\chi}$, we get 
\begin{equation}\label{achi}
2 \chi \; \frac{d A}{d \chi} = -1 
+ {\rm sign} (\dot{\chi}) \; \sqrt{K} \; , \; \; \; \; 
K \equiv \frac{\Omega}{3} \; 
\left( 1 + \frac{2 \rho_0 \chi e^{3 (1 - \gamma) A}}
{(\sigma (t) + c)^2} \right) \; .  
\end{equation}
Equations (\ref{chid}) - (\ref{achi}) will prove useful 
in reading off the essential features of the evolution. 

{\bf 3.}
We need to determine whether 
the evolution of the universe in the present model is 
singular or not. Directly obtaining explicit solutions 
to (\ref{ad})-(\ref{rho}), with $\Omega$ arbitrary, 
is not possible under general 
conditions pertaining to the observed universe. (For 
special cases, see \cite{barrow}.) Nevertheless, by 
a careful analysis, it turns out to be possible to 
read off all the features of the general solutions. We 
will show this below. Particularly, we will show that 
a homogeneous isotropic universe, such as our observed 
one, evolves in the present model with no big bang or 
any other singularity. 

First, an observation. For the singularities to be 
absent, {\em all} curvature invariants must be finite. 
A sufficient condition for this, proved in the Appendix, 
is that the quantities in (\ref{qty}) be all finite. 
However, some or all of these quantities can potentially 
diverge when $e^A$ and/or $\chi$ vanish or reach 
a critical value, or when $\Omega \to \infty$. Hence, 
we pay particular attention to these special points. 

To proceed, we need initial conditions. We start with 
an initial time $t_i$, corresponding to a temperature 
$\stackrel{>}{_\sim} 10^{16}$ GeV such that GUT symmetry 
breaking, inflation, and other (matter) model dependent 
phenomena may occur for $t > t_i$ only. The initial 
conditions at $t_i$, relevant to our universe, are 
$\dot{A} (t_i) > 0$ and $\dot{\chi} (t_i) > 0$ which  
correspond to an expanding universe and a decreasing 
effective Newton's constant ($ = \frac{1}{16 \pi \chi}$). 
The choice of $t_i$ means that, for 
$t \le t_i$, the universe is dominated by ``matter'' 
with $\gamma \ge \frac{1}{3}$ for as long as the scale 
factor $e^A$ is decreasing. We will first analyse 
the evolution for $t < t_i$, which is when the big bang 
singularity arises in GR, and then for $t > t_i$. 

{\bf 3a. $t \le t_i$ :} 
Now, $\dot{A} (t_i)$ and $\dot{\chi} (t_i)$ 
and, hence, the constant $c$ in (\ref{chid}) and 
$\frac{d A}{d \chi} (t_i)$ are positive. 
Therefore, $K (t_i) > 1$ in (\ref{achi}). As $t$ 
decreases below $t_i$, the quantities 
in $K$ vary as follows: $A$ decreases, since 
$\dot{A} > 0$. The integral in (\ref{chid}) is negative 
since $t < t_i$; its prefactor is $\le 0$, since $A$ 
decreases and therefore $\gamma \ge \frac{1}{3}$. 
Therefore $\sigma (t)$ is non decreasing. Equation 
(\ref{chid}) then implies, since $A$ decreases, that 
$\dot{\chi} (t) > \dot{\chi} (t_i) > 0$. Hence, $\chi$ 
decreases. In turn, $\Omega$ decreases as $\chi$ 
decreases, which follows from the properties of 
$\Omega$. The net consequence, therefore, is that $K$ 
decreases monotonically as $t$ decreases below $t_i$. 

Note that the lowest value of $K$ is $\frac{\Omega_0}{3} \le 
\frac{1}{9}$, achievable at $\chi = 0$. Since $K (t_i) > 1$ 
initially, it follows from the above discussion that 
there exists a time $t_m < t_i$ where $K (t_m) = 1$ with 
$\chi (t_m) > 0$. Then, $\frac{d A}{d \chi} (t_m) = 0$ 
implying, since $\dot{\chi} (t_m) > 0$, that 
$\dot{A} (t_m) = 0$. Furthermore, since $K (t)$ remains 
finite and $\chi (t)$ remains non zero for 
$t_m \le t \le t_i$, it follows from (\ref{achi}) that 
$A (t_m) = {\rm finite}$, implying that the scale factor 
$e^{A (t_m)}$ is non zero. Thus, for $t < t_i$, 
the universe continues to shrink and reaches a minimum 
non zero size at $t = t_m$. (That this critical point of 
$A$ is a minimum is physically obvious; it can also 
be seen from the positivity of $\ddot{A} (t_m)$.) 
The precise values of $t_m$ and of $A, \; \chi$, and 
$\Omega$ at $t_m$ are model dependent and nothing further 
can be said about them, except that $\Omega (t_m) \le 3$  
(equality iff $\rho_0 = 0$) since $K (t_m) = 1$. 

However, the above information suffices for our purposes. 
For $t_m \le t \le t_i$, $\chi$ and $e^A$ remain non zero. 
Hence, it follows from (\ref{chid}) that 
$\dot{\chi} (t)$ is finite. Therefore, the quantities 
in (\ref{qty}) all remain finite, implying that all 
the curvature invariants are also finite. Thus, there 
is no singularity for $t_m \le t \le t_i$ \cite{mb}. 

{\bf 3b. $\; t \le t_m$ :}
As $t$ decreases below $t_m, \dot{A}$ becomes 
negative and, hence, $A$ increases. The integral 
in (\ref{chid}) is negative as before. However, its 
prefactor can be positive, since $A$ increases now 
and therefore $\gamma$ can be $< \frac{1}{3}$. Hence, 
$\sigma (t)$ can be negative for $t < t_m$. 

First consider the alternative that 
$\sigma (t) + c > 0$ for 
all $t < t_m$. Then $\dot{\chi} > 0$, implying that 
$\chi$ continues to decrease for all $t < t_m$. 
Thus, singularities may arise as $\chi \to 0$. 
In this limit, $\Omega \to \Omega_0$. In (\ref{achi}), 
$\chi e^{3 (1 - \gamma) A}$ is, obviously, either 
$\ll 1$, or ${\cal O} (1)$, or $\gg 1$. Considering each 
of these possibilities in turn, and solving consistently 
equations (\ref{achi}) first and then (\ref{chid}), 
yields the unique solution as $\chi \to 0$: 
\begin{eqnarray}
{\rm for } \; \; \Omega_0 \ne \frac{1}{3} \; : & & 
e^{A - A_0} = \left( \frac{a t}{m} + b \right)^n 
\; , \; \; \; \; 
\chi = \chi_0 \left( \frac{a t}{m} + b \right)^m  
\label{chi0} \\ 
{\rm for } \; \; \Omega_0 = \frac{1}{3} \; : & & 
e^{A - A_0} = e^{- \frac{a t}{3}} \; , 
\; \; \; \;   \; \; \; \;   \; \; \; \;   \; \; 
\chi = \chi_0 e^{a t} \; , \label{chi0e}
\end{eqnarray}
where $A_0, \; \chi_0 > 0, \; a > 0$, and $b$ are 
(model dependent) constants, and 
\begin{equation}\label{mn}
n = \frac{3 - \sqrt{3 \Omega_0}}
{3 (1 - \sqrt{3 \Omega_0})} \; , \; \; \; \; 
m = \frac{- 2}{1 - \sqrt{3 \Omega_0}} \; . 
\end{equation}
Consistency of these solutions requires that $\gamma > 0$ 
(see below for $\gamma \le 0$) 
and $\sqrt{3 \Omega_0} > \frac{1 - 3 \gamma}{1 - \gamma}$. 
Note that $\Omega_0 = \frac{1}{3}$ is perhaps the best 
choice, which is also significant from another point of 
view, as discussed at the end. 

Let $\Omega_0 > \frac{1}{3}$. Then, $m > 0$ in (\ref{mn}). 
Since $\chi \to 0$, it follows from (\ref{chi0}) that 
$t \to - \frac{m b}{a} = {\rm finite}$, implying that 
$\chi$ vanishes at a finite time in the past. 
Also, the quantities in (\ref{qty}), for example 
$\frac{\dot{\chi}}{\chi}$, diverge, implying that 
the curvature invariants, including the Ricci scalar, 
diverge. Thus, for $\Omega_0 > \frac{1}{3}$, there is 
a singularity at a finite time in the past. 

Let $\Omega_0 \le \frac{1}{3}$, which is the case in 
our model, see (\ref{123}). Then, $m < 0$ in (\ref{mn}). 
Since $\chi \to 0$, it follows from (\ref{chi0}) 
and (\ref{chi0e}) that $t \to - \infty$.  Evaluating 
now the quantities in (\ref{qty}), it can be seen that 
they  are all finite, implying that all the curvature 
invariants are finite. Thus, for 
$\Omega_0 \le \frac{1}{3}$, there is no singularity 
at any time in the past. 

Now consider the remaining alternative that 
$\sigma (t) + c$ is not positive for all $t < t_m$ 
which will be the case, for example, for $\gamma \le 0$. 
Then, $\sigma (t) + c$ and, hence, $\dot{\chi}$ must 
vanish at some time $t = t_r$, with $\chi (t_r)$ 
non vanishing. (The case of vanishing $\chi (t_r)$ 
is similar to the one discussed above.) Clearly, as 
follows from equations (\ref{ad}), 
(\ref{chidd}), (\ref{chid}), and (\ref{achi}), all 
the quantities in (\ref{qty}) remain finite at $t_r$ 
and, hence, there is no singularity at $t = t_r$. 

For $t < t_r, \; \dot{\chi} (t) < 0$ and, hence, 
$\chi (t)$ increases. 
The evolution can then be analysed along 
similar lines as above. In particular, if $\chi$ 
remains in the range $0 < \chi < 1$ then 
no singularities arise. However, depending on 
the details of matter content and $\Omega (\chi)$, 
the values of $e^A$ and $\chi$ may oscillate back 
and forth. But, the singularities can potentially 
arise only when $\chi \to 0$, or $\chi \to 1$ where 
$\Omega \to \infty$. $\chi \to 0$ is analysed above 
and $\chi \to 1$ will be analysed presently. 

{\bf 3c. $\; t \ge t_i$ :} $\dot{A} (t_i)$ and 
$\dot{\chi} (t_i)$ are positive, and the evolution will 
now lead to the present day universe. For $t > t_i$, 
GUT symmetry breaking, inflation, and other 
(matter) model dependent phenomena may occur, but are all 
non singular. Also, in all such phenomena, both $e^A$ and 
$\chi$ increase. Thus, as $t$ increases beyond $t_i$, 
$\chi$ increases and eventually $\chi \to 1$. Consider 
now this limit, where singularities may arise because 
$\Omega \to \infty$. 

As $\chi \to 1, \; \Omega \to \Omega_1 
(1 - \chi)^{- 2 \alpha}$, see (\ref{123}). Then, to 
an excellent approximation, $\sigma (t)$ is constant and 
equation (\ref{achi}) can be solved exactly. Proceeding 
as in {\bf 3b} above yields the unique solution, which 
also describes the present day universe: 
\begin{equation}\label{chi1}
e^{A - A_0} = t^{\frac{2}{3 (1 + \gamma)}} 
\; , \; \; \; \; 
\chi = 1 - \chi_0 t^{- \frac{1 - \gamma}
{(1 + \gamma) (1 - \alpha)}} \; ,  
\end{equation}
where $A_0$ and $\chi_0 > 0$ are constants. It can now be 
easily seen that the quantities in (\ref{qty}) are all 
finite, implying that all the curvature invariants are 
finite. Thus, there is no singularity as $\chi \to 1$. 
This shows that there is no singularity for $t > t_i$ 
(and also for $t < t_r$ in {\bf 3b} above). 

Thus, in the present day universe in our model, 
$\chi \to 1$ and $\Omega \to \infty$. Also, as 
follows from (\ref{123}), $\frac{1}{\Omega^3} 
\frac{d \Omega}{d \chi} (today) \propto (1 - \chi)^{4 
\alpha - 1} \to 0$. Therefore, our model satisfies 
the observational constraints imposed by solar system 
experiments, {\em viz.} $\Omega (today) > 2000$ and 
$\frac{1}{\Omega^3} \frac{d \Omega}{d \chi} (today) 
< 0.0002$ (see \cite{will}, pg. 117, 124-5, and 339). 

{\bf 4.} 
We have thus shown that a homogeneous isotropic 
universe, such as our observed one, evolves in 
the present model with no big bang or any other 
singularity. The time continues indefinitely into 
the past and the future, without encountering 
any singularity. This is a generic result valid for 
any function $\Omega$ satisfying only the constraints 
(\ref{123}). Note that the evolutions given in 
\cite{barrow,k} for specific choices of 
$\Omega (\chi)$ and/or $\rho$ and $\gamma$ all conform 
to the evolution described here, and are singularity 
free only when $\Omega (\chi)$ satisfies (\ref{123}). 

Now, a large class of $\Omega$ satisfying (\ref{123}) 
can be easily constructed. An important question to ask, 
however, is whether such a function $\Omega$ can arise 
from string theory. We now discuss critically a few 
possibilities. 

In this context, Einstein metric formulation given in 
\cite{ein} is more natural where $\phi$ is the dilaton 
and $\chi (\phi)$ the arbitrary function. The strong 
coupling limit in string theory, relevant for 
singularities, corresponds to diverging effective 
Newton's constant ($= \frac{1}{16 \pi \chi}$) 
and, thus, to the limit $\chi (\phi) \to 0$. In this 
limit, $\phi \to - \infty$ and ${\rm ln} \chi \to 
\frac{\phi}{\sqrt{\Omega_0}}$ in our model (see 
equation (\ref{123}) and \cite{ein}). 

(1) Note that, in string theory, ${\rm ln} (\chi (\phi))$ 
can be thought of as Kahler potential for $\phi$, 
which is expected to be modified at 
strong coupling by non perturbative effects. But, 
such modifications are exponential in nature and, hence, 
are unlikely to lead to a $\chi (\phi)$ as required here. 

(2) However, at strong coupling, $\phi$ here may not 
be the stringy dilaton, but instead a combination of 
the stringy dilaton and other compactification dependent 
moduli. In that case, after writing the relevant effective 
action as given in \cite{ein}, the Kahler potential 
${\rm ln} \chi$ for this `effective dilaton' $\phi$ 
may well turn out to be of the required form. 

(3) Another, perhaps more promising, 
possibility is the following which 
arises when $\Omega_0 = \frac{1}{3}$ (see equation 
(\ref{123}) and \cite{ein}). This case corresponds to 
a five dimensional space time compactified to four 
dimensions on a circle \cite{gm}. In recent developments 
in string theory at strong coupling, analogous phenomena 
relating a $d$-dimensional theory and 
a $(d + 1)$-dimensional theory on a circle are found to 
occur: Using S-duality symmetries, Witten has discovered 
that the ten dimensional string at strong coupling is 
an eleven dimensional (M-)theory compactified on a circle 
\cite{w,rev}. A similar phenomenon, 
that the three dimensional string at strong coupling 
is a four dimensional theory compactified on a circle, 
is at the heart of Witten's novel proposal for solving 
the cosmological constant problem \cite{cc}; for its 
possible stringy realisation see \cite{v}. A similar 
phenomenon in four/five dimensions, if exists, 
is likely to lead to a $\chi (\phi)$ as required here. 

Admittedly, at present, these are plausibility arguments 
only. Nevertheless, given the elegant way the present 
model resolves the big bang singularity it is worthwhile 
to derive it, perhaps along the above lines, from 
a fundamental theory such as string theory. On the other 
hand, however, such a model, even if not derivable from 
string theory, can stand on its own as 
an interesting model for singularity free cosmology. One 
may then study its implications in other cosmological and 
astrophysical contexts, which are likely to be novel 
\cite{k}. Such studies are fruitful and are likely 
to provide valuable insights. 

{\bf Acknowledgement:} We thank the referee for pointing 
out \cite{lessner}. 

\vspace{4ex}

{\bf Appendix:}
Curvature invariants are constructed using metric tensor, 
Riemann tensor, and covariant derivatives. When the metric 
is diagonal, every term in any curvature invariant can be 
grouped into three types of factors: 
(A) $\sqrt{g^{\mu \mu} g^{\nu \nu} g^{\lambda \lambda} 
g^{\tau \tau}} \; R_{\mu \nu \lambda \tau}$, 
(B) $\sqrt{g^{\mu \mu } g^{\nu \nu } g_{\lambda \lambda}} 
\; \Gamma_{\mu \nu}^\lambda$, and 
(C) $\sqrt{g^{\mu \mu}} \; \partial_\mu$ acting 
multiply on (A) and (B) type factors (no summation over 
repeated indices). 

Using the metric $d s^2 = - d t^2 + e^{2 A(t)} \; 
(\frac{d r^2}{1 - k r^2} + r^2 d S_2^2)$ and evaluating 
explicitly the above factors, one obtains that (A) and 
(B) type factors are functions of $\ddot{A}, \dot{A}$, 
and $e^{- A}$. The action of (C) produces extra time 
derivatives. Thus, it follows that the curvature 
invariants are functions of $e^{- A}$ and 
$\frac{d^n A}{d t^n}, \; n \ge 1$. 

By repeated use of equations (\ref{ad})-(\ref{rho}), 
$\frac{d^n A}{d t^n}$ for any $n \ge 1$ and, hence, 
any curvature invariant, can be expressed in terms of 
the following quantities: 
\begin{equation}\label{qty}
e^{- A}; \; \; \; 
\frac{\rho}{\chi}, \;  
\frac{\rho}{\chi \Omega}; \; \; \; 
\frac{\dot{\chi}}{\chi}, \; 
\frac{\Omega \dot{\chi}^2}{\chi^2}; \; \; \; 
{\rm and} \; \; \; 
\frac{\chi^n}{\Omega} \; \frac{d^n \Omega}{d \chi^n} \; 
\left( \frac{\dot{\chi}}{\chi} \right)^n, \; \; 
\forall n \ge 1 \; . 
\end{equation}
The resultant expressions are finite if the above 
quantities are finite. The algebra is straightforward 
and, hence, we omit the details here. 

Therefore, {\em a sufficient condition for all the curvature 
invariants to be finite is that the quantities in 
(\ref{qty}) be all finite}.

\end{document}